\documentclass[twocolumn,showpacs,amsmath,amssymb]{revtex4}
\usepackage{amssymb}
\usepackage{txfonts}
\usepackage{bbm}
\usepackage{graphicx}
\usepackage{appendix}
\usepackage{epsf}
\usepackage{epstopdf}

\begin{document}

\title{Consistent picture for the electronic structure around a vortex core in iron-based superconductors}

\author
{Tao Zhou$^{1,2}$, Z. D. Wang$^{2}$, Yi Gao$^{3}$, and C. S. Ting$^{3}$}
\affiliation{$^{1}$ College of Science, Nanjing University of Aeronautics and Astronautics, Nanjing 210016, China\\
$^{2}$ Department of Physics and Center of Theoretical and Computational Physics,
The University of Hong Kong, Pokfulam Road,Hong Kong China\\
$^{3}$Texas Center for Superconductivity and Department of
Physics, University of Houston, Houston, Texas 77204, USA
}

\date{\today}
\begin{abstract}

Based on a two-orbital model and taking into account the presence of the impurity,
we studied theoretically the electronic structure in the vortex core of the iron-Pnictide superconducting materials.
The vortex is pinned when the impurity is close to the vortex core.
The bound states shows up for the unpinned vortex and are wiped out by a impurity. Our results are in good agreement with recent experiments
and present a consistent explanation for the different electronic structure of vortex core revealed by experiments on different materials.
\end{abstract}
 \maketitle

The iron-based superconducting (SC) materials have
attracted much attention since their discovery~\cite{kam}. Similar to the cuprate  materials,
the superconductivity in iron-pnictides can be achieved
by doping either holes or electrons into the parent compound, among which the compound BaFe$_2$As$_2$, named as "122" system,
is one of the most studied iron-based systems. The superconductivity can be realized by substituting the Barium ions by Kalium ions or iron ions by the cobalt ions,
corresponding to the hole-doped materials Ba$_{1-x}$K$_x$Fe$_2$As$_2$ and electron-doped materials BaFe$_{2-x}$Co$_x$As$_2$, respectively.

The scanning tunneling microscope (STM) experiment has been a powerful tool to investigate the electronic structure of SC materials.
Moreover, it can reveal the vortex core structure and possible vortex state in presence of the magnetic field. Recently, the STM experiments were performed on both BaFe$_{2-x}$Co$_x$As$_2$~\cite{hof} and Ba$_{1-x}$K$_x$Fe$_2$As$_2$~\cite{shan} compounds. However, the measurements on these two doped materials revealed significantly different results.
For the hole-doped compound Ba$_{1-x}$K$_x$Fe$_2$As$_2$, the vortex forms a regular distorted triangle lattice. The strong bound state at the vortex core center was observed clearly with the energy peaked at the negative energy. The intensity decreases and the peak splits with a dominant weight at negative energy when away from the center and finally evolves to the gap edges.
 However, for the compound BaFe$_{2-x}$Co$_x$As$_2$, different results were reported, namely, the vortex forms disordered lattice and no bound state was observed at all.

The contrast experimental observations in Ba$_{1-x}$K$_x$Fe$_2$As$_2$ and BaFe$_{2-x}$Co$_x$As$_2$ are intriguing.
Theoretically, the existence of bound states is robust and can be reproduced based on various effective models for iron-pnictides,
while the position of the resonance peak is model dependent~\cite{hu,jiang,gao,wang}. Especially, the negative bias peak is obtained~\cite{gao,wang} based on a two-orbital model proposed in Ref.~\cite{zhang}, a three-orbital model in Ref.~\cite{dag}, and a five-orbital one in Ref.~\cite{kur}.
 Although there has been no theoretical calculation to explain the absence of the bound states in BaFe$_{2-x}$Co$_x$As$_2$, a possible explanation is proposed in Refs.~\cite{hu,gao}, i.e., the strong spin-density-wave (SDW) induced by the magnetic field inside the core would suppress the bound state. While the in-gap peak structure is still visible in the numerical results even in presence of the strong SDW order~\cite{gao}. We note that for Cobalt doped samples, the static SDW order disappears at the doping around $\delta=0.065$~\cite{lap,les} and the STM experiment was performed at the doping density $\delta=0.1$~\cite{hof}, thus there is no SDW order in the bulk. The SDW fluctuation inside the vortex could be induced by the magnetic field and we expect it may suppress the bound state more or less while it is not strong enough to wipe out the bound states completely.
On the other hand, this issue was also discussed qualitatively in Ref.~\cite{shan} when comparing their experimental results with Ref.~\cite{hof}.
It was proposed that the in-plane Cobalt ions or multi-band nature of the iron pnictides may account for such discrepancy. Since the Ba$_{1-x}$K$_x$Fe$_2$As$_2$ and BaFe$_{2-x}$Co$_x$As$_2$ share the same parent compound,  they should have similar multi-band structures. Thus we expect that the Cobalt ions, which may serve not only as dopants but also impurities, should be responsible for this discrepant behavior. In fact,  the impurity effect and its role in the uniform SC state in iron-pnictides have been studied intensively~\cite{zhang,chub,bang,park,seng,voro,zhou,tsai} and it was proposed that it can be used to probe the pairing symmetry~\cite{zhou,zhang,tsai}. In presence of the magnetic field, the interplay between the impurity and vortex is of interest and studied in cuprates~\cite{zhu,kim}, while the relevant study about the iron-pnictides is still awaited. Therefore, it is timely and interesting to address the impurity effect in presence of the field and present a consistent picture for the STM results for Co-doped and K-doped materials.

 In this paper, we study the single impurity effect
based on a two-orbital model~\cite{zhang} which has given a sound explanation for the vortex bound state in Ba$_{1-x}$K$_x$Fe$_2$As$_2$~\cite{gao}.
The model is given by
\begin{eqnarray}
H=&-\sum_{{\bf i}\mu {\bf j}\nu\sigma}(t_{{\bf i}\mu
{\bf j}\nu}c^\dagger_{{\bf i}\mu\sigma}c_{{\bf j}\nu\sigma}+h.c.)-\mu\sum_{{\bf i}\mu\sigma}c^{\dagger}_{{\bf i}\mu\sigma}c_{{\bf i}\mu\sigma}
\nonumber\\& +\sum_{{\bf i}\mu {\bf j}\nu\sigma}(\Delta_{{\bf i}\mu
{\bf j}\nu}c^\dagger_{{\bf i}\mu\sigma}c^{\dagger}_{{\bf j}\nu\bar{\sigma}}+h.c.)+\nonumber\\&
U\sum_{{\bf i}\mu\sigma\neq\bar{\sigma}} \langle
n_{{\bf i}\mu\bar{\sigma}}\rangle
n_{{\bf i}\mu\sigma}+U^{\prime}\sum_{{\bf i},\mu\neq\nu,\sigma\neq\bar{\sigma}}\langle
n_{{\bf i}\mu\bar{\sigma}}\rangle n_{{\bf i}\nu{\sigma}}
+\nonumber\\
&+(U^{\prime}-J_H)\sum_{{\bf i},\mu\neq\nu,\sigma}\langle
n_{{\bf i}\mu\sigma}\rangle n_{{\bf i}\nu\sigma}+
\sum_{{\bf i_m}\mu\sigma}V_s c^{\dagger}_{{\bf
i_m}\mu\sigma}c_{{\bf i_m}\mu\sigma}\;,
\end{eqnarray}
where ${\bf i}=(i_x,i_y)$, ${\bf j}=(j_x,j_y)$ are the site indices,
$\mu,\nu=1,2$ are the orbital indices, and $\mu$ is the chemical
potential. $n_{{\bf i}\mu\sigma}$ is the density operator at site ${\bf
i}$ and orbital $\mu$, with spin $\sigma$. The quantity $U^{\prime}$
is taken to be $U-2J_H$. $V_s$ is the impurity potential.
The impurities are located at the site $i_m$. Generally speaking, for the magnetic impurity, the impurity scattering would include both the potential scattering and the magnetic part. Based on first-principles
electronic-structure calculations, the potential scattering potential is much stronger than the magnetic scattering one~\cite{kem}.
In the present work,
the Cobalt irons are taken as a strong potential scattering center, while the weaker magnetic part is not considered.

The hopping constants 
are written as
$t_{{\bf i}\mu
{\bf j}\nu}=t_{{\bf i}\mu
{\bf j}\nu,0}\exp[i(\pi/\phi_0)]\int^{\bf r_j}_{\bf r_i} {\bf A}({\bf r}) d{\bf r}$ in presence of the magnetic field, with
$\Phi_0$ is the superconducting flux quantum and ${\bf A}=(-By,0,0)$ is the potential in the Laudau gauge.
Following Ref.~\cite{zhang}, we use
\begin{eqnarray}
t_{{\bf i}\mu,{\bf i}\pm\hat{\alpha}\mu}&=&t_1 \qquad (\alpha=\hat{x},\hat{y})\;,\\
t_{{\bf i}\mu,{\bf i}\pm(\hat{x}+\hat{y})\mu}&=&\frac{1+(-1)^{\bf i}}{2}t_2+\frac{1-(-1)^{\bf i}}{2}t_3\;,\\
t_{{\bf i}\mu,{\bf i}\pm(\hat{x}-\hat{y})\mu}&=&\frac{1+(-1)^{\bf i}}{2}t_3+\frac{1-(-1)^{\bf i}}{2}t_2\;,\\
t_{i\mu,{\bf i}\pm\hat{x}\pm\hat{y}{\nu}}&=&t_4 \qquad (\mu\neq \nu)\;.
\end{eqnarray}

The mean-field Hamiltonian can be diagonalized by solving the  Bogoliubov-de Gennes (BdG) equations,
\begin{equation}
\sum_{\bf j}\sum_\nu \left( \begin{array}{cc}
 H_{{\bf i}\mu {\bf j}\nu\sigma} & \Delta_{{\bf i}\mu {\bf j}\nu}  \\
 \Delta^{*}_{{\bf i}\mu {\bf j}\nu} & -H^{*}_{{\bf i}\mu {\bf j}\nu\bar{\sigma}}
\end{array}
\right) \left( \begin{array}{c}
u^{n}_{{\bf j}\nu\sigma}\\v^{n}_{{\bf j}\nu\bar{\sigma}}
\end{array}
\right) =E_n \left( \begin{array}{c}
u^{n}_{{\bf i}\mu\sigma}\\v^{n}_{{\bf i}\mu\bar{\sigma}}
\end{array}
\right),
\end{equation}
with
\begin{eqnarray}
H_{{\bf i}\mu {\bf j}\nu\sigma}=&&-t_{{\bf i}\mu {\bf
j}\nu}+[U\langle n_{{\bf i}\mu\bar{\sigma}}\rangle+(U-2J_H)\langle
n_{{\bf i}\bar{\mu}\bar{\sigma}}\rangle\nonumber
\\&&+(U-3J_H)\langle n_{{\bf i}\bar{\mu}\sigma}\rangle+v_s
\delta_{{\bf i},{\bf i_m}}-t_0]\delta_{\bf ij}\delta_{\mu\nu}\;,
\end{eqnarray}
and
\begin{eqnarray}
\Delta_{{\bf i}\mu {\bf j}\nu}=\frac{V_{{\bf i}\mu {\bf j}\nu}}{4}\sum_n
(u^{n}_{{\bf i}\mu\uparrow}v^{n*}_{{\bf j}\nu\downarrow}+u^{n}_{{\bf j}\nu\uparrow}v^{n*}_{{\bf i}\mu\downarrow})\tanh
(\frac{E_n}{2K_B T})\;,
\end{eqnarray}
\begin{eqnarray}
\langle n_{{\bf i}\mu}\rangle &=&\sum_n
|u^{n}_{{\bf i}\mu\uparrow}|^{2}f(E_n)+\sum_n
|v^{n}_{{\bf i}\mu\downarrow}|^{2}[1-f(E_n)]\;.
\end{eqnarray}
Here $V_{{\bf i}\mu {\bf j}\nu}$ is the pairing strength and $f(x)$
is the Fermi-Dirac distribution function.

The LDOS is calculated according to
\begin{equation}
\rho_{\bf i}(\omega)=\sum_{n\mu}[|u^{n}_{{\bf i}\mu\sigma}|^{2}\delta(E_n-\omega)+
|v^{n}_{{\bf i}\mu\bar{\sigma}}|^{2}\delta(E_n+\omega)]\;,
\end{equation}
where the delta function $\delta(x)$ is taken as
$\Gamma/\pi(x^2+\Gamma^2)$, with the quasiparticle damping
$\Gamma=0.01$.

In the numerical calculations, following previous studies~\cite{zhou1,gao1,zhou2}, we set $t_{1-4}=1, 0.4, -2, 0.04$. The
on-site Coulombic interaction $U$ and Hund's coupling $J_H$ are
taken as $3.4$ and $1.3$, respectively. The pairing is chosen as next-nearest-neighbor
intra-orbital pairing with the pairing strength $V=1.2$, which will reproduce the $s_{x^2y^2}$-wave pairing symmetry, being consistent the experimental~\cite{ding,ctchen} and theoretical proposal~\cite{mazin,kaz,fwang,yao} for the gap symmetry in iron pnictides.
The doping density is fixed at the electron-doped $\delta=0.1$ according to the experimental value~\cite{hof}.
A $40\times 20$ square lattice with the periodic
boundary conditions is used and a $10\times 20$ super-cell is taken to calculate
the LDOS.

\begin{figure}
\centering
  \includegraphics[width=8cm]{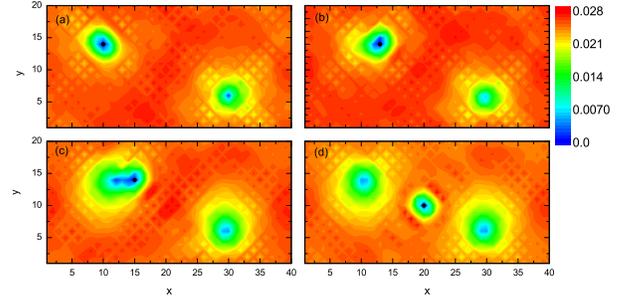}
\caption{(Color online) The intensity plots of the gap magnitude  in presence of one impurity  at the site $(10,14)$, $(13,14)$, $(16,14)$ and $(20,10)$, respectively. The impurity locations are marked by the black spots.}
\end{figure}

 The magnitudes of the SC gap with a strong impurity $V_s=100$ at different locations are plotted in Fig. 1.
 Without the impurity, the vortex cores
would appear at the site $(10,14)$ and $(30,6)$ (not presented here). When the impurity is put at the core center or near the vortex core, the vortex would be pinned by the impurity, as seen in Figs.1(a) and 1(b). Here the pinned effect in the present two-orbital model is similar to previous results in cuprates~\cite{zhu,kim}.
The critical distances $r_{c}$ between the
impurity and the core center
(in a clean system) can be defined below which the vortex core center will be dragged to the impurity site.
 For the present parameters $r_c$ is about four lattice constant. It is insensitive to the direction of the impurity but sensitive to the impurity potentials and electron interaction, which is not concerned with in the present work. As the distance is larger than the critical one, as seen in Fig.1(c), the vortex and the impurity will be separated. Since the distance is just slightly larger than the critical one, for this case the vortex is still dragged to be close to the impurity site.
The above pinned and dragged effects
present a reasonable explanation why the vortex lattice is disordered in Co-doped materials~\cite{hof}. As the distance increases further, as seen in Fig.1(d), the impurity has no effect on the vortex.

\begin{figure}
\centering
  \includegraphics[width=8cm]{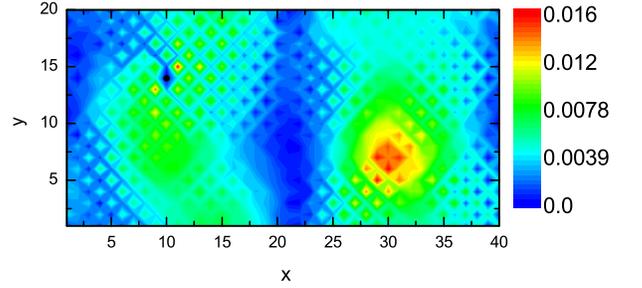}
\caption{(Color online) The intensity plot of the magnetic order in presence of one impurity  at the site $(10,14)$.}
\end{figure}

The intensity plot of the magnetic order (defined as $|1/2 \sum_\mu (n_{i\mu\uparrow}-n_{i\mu\downarrow})|$) is plotted in Fig.2. As seen,
the magnetism is induced by the field and reaches the maxima value near the unpinned vortex core center. Away from the vortex, the magnetism disappears gradually. These features are qualitatively the same with previous studies about field induced SDW order without considering the impurity effect~\cite{hu,jiang,gao}.
For the case of pinned vortex, the magnetic order will be suppressed significantly near the impurity site and such suppression effect is similar to previous studies about the impurity effect in the SDW state~\cite{zhou2}. It is also  notable that for a unpinned vortex, the field-induced magnetic order is weak for the present parameters, and thus the suppression of the low-energy bound states is also expected to be quite weak.
Since we used the same model and the same parameters, we actually  have a good consistent explanation for experimental results in
BaFe$_{2-x}$Co$_x$As$_2$~\cite{zhou1,gao1}.
Therefore,
we conclude that the field-induced SDW cannot account for the disappearance of the bound states in Co-doped materials.

\begin{figure}
\centering
  \includegraphics[width=8cm]{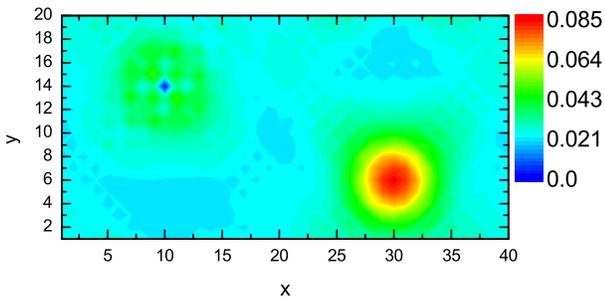}
\caption{(Color online) The averaged low energy ($\int^{\epsilon}_{-\epsilon} \rho(\omega)d\omega$ with $\epsilon=0.05$) LDOS map with one impurity at $(10,14)$.}
\end{figure}

We now study the low energy electronic structure for the pinned and unpinned vortex.
The intensity plot of the LDOS at the real space with one pinned vortex and one unpinned vortex is shown in Fig.3.
As is seen, near the unpinned vortex the low-energy LDOS is enhanced and largest at the core center, indicating the existence of the low-energy bound state in the vortex core, agreeing qualitatively with the numerical results for hole-doped materials~\cite{gao}.
While for the pinned vortex with a strong impurity locating at the vortex center, the LDOS at the impurity site is zero. Near the center the spectra is suppressed significantly, and it seems that no bound state exists from the whole region of the pinned vortex.

\begin{figure}
\centering
  \includegraphics[width=8cm]{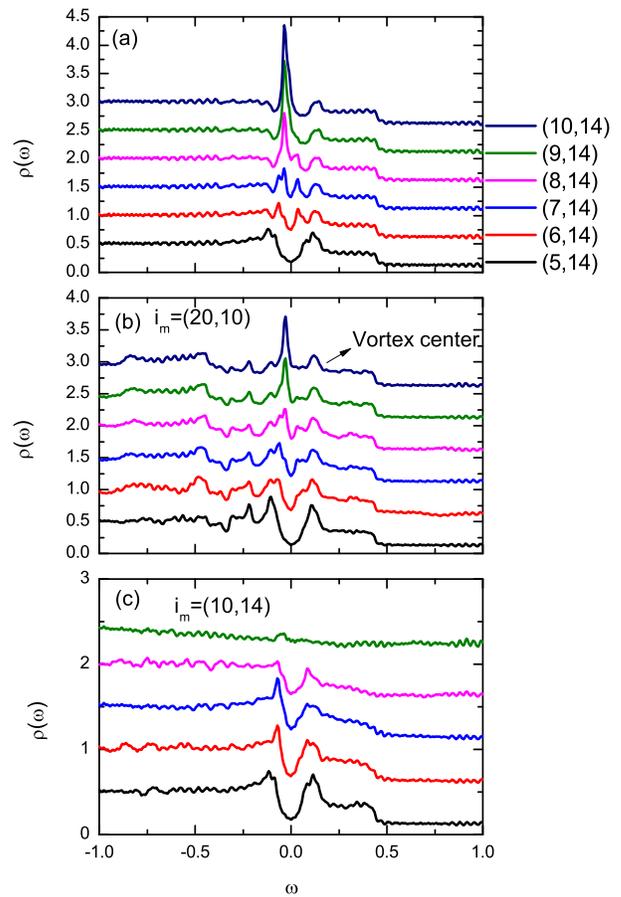}
\caption{(Color online) The LDOS spectra for from the vortex center [the site $(10,14)$] to the bulk respectively.
Panel (a) is for the clean system and panels (b) and (c) are for the cases of one impurity with
$i_m$ the impurity location.  }
\end{figure}

The LDOS spectra from the bulk to the vortex center
 are plotted in Fig.4. The spectra without impurity are shown in Fig.4(a). As is seen, there is a strong in-gap peak at the negative energy for the spectrum at the vortex core center. The peak will spilt with a stronger peak at a negative energy when away from the center. The peak intensity decreases when further away from the vortex and finally the LDOS involves to the bulk one.
 All of the features are qualitatively the same with previous results for hole-doped samples~\cite{gao}, indicating that the changing chemical potential are not important to the low energy electronic structure. The LDOS for the unpinned vortex with the impurity at the site $(20,10)$ is plotted in Fig.4(b). The spectra are similar to those in clean systems, while the intensities of the resonance peak decrease. And some features outside the SC gap appear, which are not a concern of the present work. The LDOS spectra for a pinned vortex with a impurity at $(10,24)$ are shown in Fig.4(c).
 For this potential the LDOS at the impurity site is zero (not shown here). Near the vortex core center, the LDOS is also suppressed signicantly
 and no low energy bound state exists for all the spectra we presented. This result is consistent with the STM experimental result reported in Co-doped materials~\cite{hof}. Actually, when away from the core center and when the SC coherent peak shows up, the coherent peak is enhanced at the negative energy. This is due to the impurity effect and discussed previously for the uniform SC state~\cite{zhang} and SDW state~\cite{zhou2} based on the same model.
 Since the enhancement is not significant enough and is near the gap edge, it may be covered up by the SC coherent peak. As a result, it may be difficult to be revealed by experiments.

 The LDOS spectra with different impurity potentials are shown in Fig.5. For all the potentials considered,
 the low energy states are suppressed significantly. For the positive potentials, some out-gap features show up as the potential decreases, as seen in Figs.5(a) and 5(b). For the moderate strength negative potentials $(V_s=-5)$, the bound states induced by the impurity is clearer at the negative gap edge. While such bound states reduce for stronger potentials $(V_s=-8)$. The spectra for $V_s=-100$ (not shown here) are almost the same as the spectra $V_s=+100$.

We here have given a reasonable explanation for the STM experiments on BaFe$_{2-x}$Co$_x$As$_2$. Since the experiment is for the sample $\delta=0.1$, the impurity concentration is much larger than the vortex density. For a isolated vortex and considering the critical distance is insensitive to the direction,
we can roughly estimate the pinned ratio as $(1-0.9^{\pi r_{c1}^2})$. The ratio is $99.5\%$ for $r_c=4$ and $95\%$ for $r_c=3$.
Thus it is understandable that the bound state cannot be observed in experiments.

\begin{figure}
\centering
  \includegraphics[width=8cm]{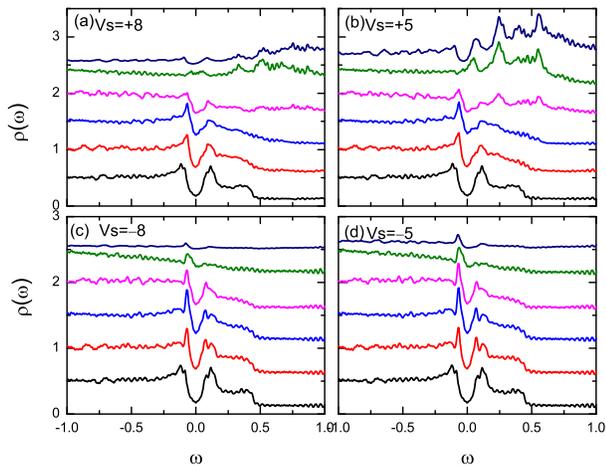}
\caption{(Color online) Similar to Fig.4 but with different impurity potentials and $i_m=(10,14)$.}
\end{figure}

In summary, based on a two-orbital model
and taking into account the impurity effect, we have studied the vortex states for iron-based superconducting materials .
The vortex will be pinned as the impurity is close to the vortex core.
The low-energy bound states
show up for a clean system and could be suppressed by the impurity. Therefore, we have presented a consistent picture for recent STM experiments on the
  Ba$_{1-x}$K$_x$Fe$_2$As$_2$ and BaFe$_{2-x}$Co$_x$As$_2$ and elaborated that
  the presence of the impurity accounts for the absence of the bound states in the vortex core
for the BaFe$_{2-x}$Co$_x$As$_2$ compound.

This work was supported by the NSFC under the Grant No. 11004105, the RGC of Hong Kong under the Nos. HKU7044/08P and HKU7055/09P and a CRF of Hong Kong. Y.G. and C.S.T. were supported by the Texas Center for Superconductivity at
the University of Houston and by the Robert A. Welch Foundation
under the Grant No. E-1146 and E-1070.

\end{document}